\documentclass[fleqn,usenatbib,referee]{mnras}
\usepackage{newtxtext,newtxmath}

\usepackage[T1]{fontenc}
\usepackage{ae,aecompl}

\usepackage{graphicx}	
\usepackage{amsmath}	
\usepackage{amssymb}	
\usepackage{subfigure}	
\usepackage{array}
\usepackage{booktabs} 
\usepackage{threeparttable}
\usepackage{textcomp,booktabs}
\usepackage[usenames,dvipsnames]{color}
\usepackage{colortbl}
\definecolor{mygray}{gray}{.9}
\definecolor{mypink}{gray}{.5}
\usepackage{soul}
\usepackage{numprint} 

\DeclareMathOperator{\inner}{in}                          
\DeclareMathOperator*{\ISCO}{ISCO} 

\title[Reflection study of MAXI J1836-194]{A detailed study on the reflection component for the Black Hole Candidate MAXI J1836-194}

\author[Yanting Dong et al.]{
Yanting~Dong,$^{1,2}$\thanks{E-mail: ytdong@nao.cas.cn}
Javier~A.~Garc\'ia,$^{3,4}$
Zhu~Liu,$^{1}$
Xueshan~Zhao,$^{1,2}$
Xueying~Zheng,$^{1,2}$\newauthor%
Lijun~Gou$^{1,2}$\thanks{E-mail: lgou@nao.cas.cn}
\\
$^{1}$The National Astronomical Observatories, Chinese Academy of Sciences, Beijing, 100101, China\\
$^{2}$University of Chinese Academy of Sciences, No.19(A) Yuquan Road, Shijingshan District, Beijing, 100049, China\\
$^{3}$Cahill Centre for Astronomy and Astrophysics, California Institute of Technology, Pasadena, CA 91125, USA\\
$^{4}$Dr. Karl Remeis-Observatory and Erlangen Centre for Astroparticle Physics, Sternwartstr.~7, 96049 Bamberg, Germany
}

\date{Accepted XXX. Received YYY; in original form ZZZ}

\pubyear{2015}

\begin{document}
\label{firstpage}
\pagerange{\pageref{firstpage}--\pageref{lastpage}}
\maketitle

\begin{abstract}
We present a detailed spectral analysis of the black hole candidate MAXI J1836-194. The source was caught in the intermediate state during its 2011 outburst by \emph{Suzaku} and \emph{RXTE}. 
We jointly fit the X-ray data from these two missions using the {\tt relxill} model to study the reflection component, and a steep inner emissivity profile indicating a compact corona as the primary source is required in order to achieve a good fit.
In addition, a reflection model with a lamp-post configuration ({\tt relxilllp}), which is normally invoked to explain the steep emissivity profile, gives a worse fit and is excluded at 99\% confidence level compared to {\tt relxill}.
We also explore the effect of the ionization gradient on the emissivity profile by fitting the data with two relativistic reflection components, and it is found that the inner emissivity flattens. These results may indicate that the ionization state of the disc is not constant. All the models above require a supersolar iron abundance higher than $\sim4.5$. However, we find that the high-density version of {\tt reflionx} can describe the same spectra even with solar iron abundance well. A moderate rotating black hole ($a_*$ = 0.84-0.94) is consistently obtained by our models, which is in agreement with previously reported values.
\end{abstract}




\section{Introduction}
\label{sec:Int}

Galactic X-ray binaries are believed to be powered by accretion onto stellar-mass black holes or neutron stars. The gases in accretion disc emit thermal radiation in UV/X-ray band \citep{sha1973}. Some fraction of the thermal photons from the disc are then inverse Compton scattered by energetic electrons in the hypothetically hot corona, producing a hard X-ray spectrum in the form of power-law, i.e. $N(E)\propto E^{-\Gamma}$.
A fraction of the high-energy photons will irradiate the cold accretion disc, generating the so-called X-ray reflection component \citep{fab1989}. The main features of the reflection spectrum are the fluorescent Fe K$\alpha$ emission line at energies of 6.4-6.97 keV (depend upon the ionization state of the disc) and the Compton hump at 20-30 keV \citep{you1999}. 

The profile of the reflection spectrum will be smeared due to the effects of Doppler shift, special relativity, and general relativity if it comes from the inner region of the accretion disc \citep{fab2000}. Observationally, the most prominent effect is that the intrinsically narrow Fe K$\alpha$ line is broadened and skewed to an asymmetric shape. The profile of the broad line, especially the red wing of the line, is directly linked to the inner radius of the accretion disc which is thought to be at the innermost stable circular orbit (ISCO), i.e. $R_{\inner}$ = $R_{\ISCO}$. Thus, by modelling the broad iron line, we can deduce the spin of the black hole based on the relation between the spin and the ISCO \citep{bar1972}. However, the line profile is readily affected by the subtraction of continuum and other components. 
Therefore, \citet{rey2013} pointed out that a more accurate measurement of the spin can be achieved by modelling the full reflection spectrum. The spin is one of the two key parameters to make a full description of a black hole. So far, we have measured dozens of stellar-mass black hole spins via X-ray reflection fitting method \citep{wan2017,bre2006, wal2013, wal2019, loh2012, bre2011, tri2019, wal2016, tom2018, mil2013, gar2018, xu2018, gar2015}.

The geometry of the corona, which is still unclear, will effect the profile of the reflection spectrum. Particularly, it has a significant impact on the emissivity profile of the reflection spectrum. The reflection emissivity profile is described by $\epsilon(r)\propto r^{-q}$, where $q$ is the emissivity index. It is normally assumed to be a broken power-law with $q = q_{\rm{in}}$, $r = R_{\rm{br}}$, and $q = q_{\rm{out}}$, where $R_{\rm{br}}$, $q_{\rm{in}}$ and $q_{\rm{out}}$ are the break radius, the emissivity index in the inner region and in the outer region, respectively.
In most cases, the break radius is usually hard to be constrained, and the emissivity index was only assumed to be a single value fixed at 3 \citep{sha1973, nov1973, rey1997} due to limited photon statistics. Of course, a good constraint on the break radius and emissivity index is also obtained for a number of systems (both AGNs and binary systems), showing a steep inner index ($q_{\rm{in}}>3$) and small break radius ($R_{\rm{br}}<6$ $R_{\rm{g}}$\footnote{$R_\mathrm{g}$ is the gravitational radius and is defined to be $R_\mathrm{g} = GM/c^2$, where $G$ is the gravitational constant, $M$ is the mass of the black hole, and $c$ is the speed of light.}), such as 1H0419-577 \citep{jia2019}, 1H0707-495 \citep{fab2011,fab2012a}, IRAS 13224-3809 \citep{fab2018}, and Mrk 335 \citep{fab2014, wil2015}, as well as for black hole binaries, such as XTE J1752-223 \citep{gar2018}, Cyg X-1 \citep{wil2012a}, GRS 1915-105 \citep{mil2013}, and MAXI J1535-571 \citep{xu2018}. The steep emissivity profile is usually explained with an extremely compact corona locating close to the black hole, in which case a large fraction of the power-law emission will be focused towards the inner region as a result of light-bending effect \citep{min2004, dau2013}. The X-ray reflection emission profile has been successfully reproduced by lamp-post model \citep{dur2016, gar2018}.

In addition, the disc ionization state can also affect the profile of the reflection emissivity, but it has not been discussed substantially in the previous studies. The ionization state of the disc at radius $r$ is defined as $\xi(r)=4$$\pi$$F_{\rm{X}}(r)/n_e(r)$, where $F_{\rm{X}}$ is the flux of the irradiation and $n_e(r)$ is the electron density of the disc at radius $r$ \citep{fab2000}. As illustrated in \citet[][see their figure 3]{svo2012} , the strong radius dependence of $F_{\rm{X}}$ will naturally lead to the radial decrease of the disc ionization for any reasonable density profile of the disc. However, the ionization is always assumed to be constant in current reflection models. The simulations by \citet{svo2012} and \citet{kam2019} indicated that the ignorance of the ionization gradient will lead to an increase in the emissivity index.

In this work, we made a detailed study on the reflection spectrum of the stellar-mass black hole candidate MAXI J1836-194. The source was discovered as an X-ray transient by the \emph{MAXI}/GSC \citep{neg2011} and \emph{Swift}/XRT \citep{ken2011} on 2011 August 30. The coordinate of MAXI J1836-194 provided by \emph{Swift}/XRT is RA/Dec (J2000) = 278.93097/-19.32004. It was identified as a black hole candidate by studying its multi-band properties \citep{mil2011, str2011, rau2011, nak2011}. MAXI J1836-194 was active for about 3 months. However, it didn't enter a soft state which suggested that this source experienced a failed outburst.  \citet{lop2019} inferred its companion as a M2 main-sequence star or later based on its near-infrared and optical properties. Low frequency quasi-periodic oscillation (QPO) are detected with \emph{Rossi X-ray Timing Explorer} (\emph{RXTE}) during the outburst \citep{jan2018}.

MAXI J1836-194 was reported to have a spin parameter of $a_{*}=0.88 \pm0.03$ at 90\% confidence level by \citet{rei2012}, using one \emph{Suzaku} spectrum during the intermediate state from its 2011 outburst. A relativistically broadened iron line was clearly shown. \citet{rei2012} used a relativistic blurring model {\tt relconv} to convolve with the reflection model {\tt refbhb} \citep{ros2007}, in which the thermal emission of the disc and the reflection emission are included in a self-consistent way. They reported a steep broken power-law emissivity profile with $q_{\rm{in}}>7.3$, $q_{\rm{out}}$ = 3.19 $_{-0.05}^{+0.07}$  and $R_{\rm{br}}$ = 3.6 $_{-0.1}^{+0.2}$ $R_{\rm{g}}$.

In this paper, we re-analysed this \emph{Suzaku} observation along with two simultaneous spectra taken by \emph{RXTE}. A much more sophisticated reflection model, namely {\tt relxill} \citep{dau2014, gar2014a}, is used. The model {\tt relxill} is the combination of the ionized reflection produced by {\tt xillver} \citep{gar2010, gar2011, gar2013} and relativistic broadening based on {\tt relline} \citep{dau2010, dau2013}. The remarkable characteristic of this model is that the reflected flux can be calculated for each point on the disc, of which the light-bending effect is also taken into account. We explore the steep power-law index using the configurations with lamp-post and the ionization gradient, respectively, and also explore the effect of disc density on reflection spectrum.

The paper is organized as follows. We describe the observation and data reduction in Section \ref{sec:data}, including both \emph{Suzaku} and \emph{RXTE} data. We present the detailed spectral analysis and results in Section \ref{sec:results}. We make discussions and conclusions in Section \ref{sec:dis} and \ref{sec:con}, respectively.

\section{Data Selection and Reduction}
\label{sec:data}
We searched the HEASARC data archive, and found one \emph{Suzaku} and 74 \emph{RXTE} observations in total. The \emph{Suzaku} observation was carried out on 2011 September 14 (MJD 55818.43). The \emph{RXTE} observations started on 2011 August 31 (MJD 55804.46) and ended on November 30 (MJD 55895.93). \citet{fer2012} and \citet{jan2018} systematically studied MAXI J1836-194 using \emph{RXTE} observations. In our work, only 2 \emph{RXTE} observations (MJD 55818.84 and MJD 55819.16) which were simultaneously observed with \emph{Suzaku} were selected. The 2 spectra are in the intermediate state with the similar flux. Their hardness ratios (HR), defined as the count rate at the energy band of 8.6-18 keV to that at 5-8.6 keV, were calculated to be 0.65 \citep{gar2015}. We listed the information including the ObsID, the start time, the end time, the exposure time, and the count rate of adopted observations in Table \ref{tab:spec}.
\begin{table*}
	\centering
	\caption{Details of the observations}
	\begin{threeparttable}[b]
	\begin{center}
    \label{tab:spec}
    \newcommand{\tabincell}[2]{\begin{tabular}{@{}#1@{}}#2\end{tabular}}
    \footnotesize
        \begin{tabular}{cccccccc}
        \toprule
        Mission&Instrument& ObsID& MJD &  \tabincell{c}{Start Time\\(in 2011)}&\tabincell{c}{End Time\\(in 2011)}&\tabincell{c}{Exp.\\(s)}&\tabincell{c}{Count Rate$^a$\\ (cts s$^{-1}$)}\\
        \midrule
        \emph{Suzaku}&
               XIS&906003010  & 55818.43& Sep 14, 10:12:23&Sep 15, 10:50:14&19440&113.7  \\
               &PIN &       &           &&&35463&1.3 \\
        \emph{RXTE}  
              &PCA &96438010104&55818.84& Sep 14, 20:12:00&Sep 14, 20:31:44&1024&100.7  \\
              &PCA &96438010105&55819.16& Sep 15, 03:56:32&Sep 15, 05:52:48&4240&102.1  \\
        \bottomrule
        \end{tabular}
    \begin{tablenotes}
        \item[$a$] Count rate is measured in 1.2-10.0 keV, 15.0-50.0 keV, and 3.0-25.0 keV for XIS, PIN, and PCA, respectively.
    \end{tablenotes}
    \end{center}
    \end{threeparttable}
\end{table*}
\subsection{\emph{Suzaku} observation}
\label{subsec:suzaku}
There are one X-ray Imaging Spectrometer (XIS, \citealt{koy2007}) and one Hard X-ray detector (HXD, \citealt{tak2007}) onboard \emph{Suzaku}. The XIS consists of four CCDs. Since one of them, XIS2, had broken down in 2006 November, the remaining three, namely XIS0, XIS1 and XIS3, were operated in the ``0.5s burst mode''. The window size, and editing mode were ``1/4 window, $3\times3$/$5\times5$'', respectively. The HXD consisting of Si PIN photo-diodes and GSO scintillation counters was operated in the ``normal'' mode. The observation was performed at XIS nominal position. Following \citet{rei2012}, we only analysed the XIS0, XIS3 and PIN data. The archival data has been reprocessed and rescreened using the \emph{Suzaku} pipeline (version 3.0.22.44) with the calibration database hxd20110913, xis20160607, and xrt20110630. The latest calibration products of XIS on 2018 August 23 only improved the redistribution matrices around the Si-K edge which will be ignored in our analysis. We generated the cleaned event files with HEASOFT version 6.19 following the \emph{Suzaku} data analysis guide{\footnote{https://heasarc.gsfc.nasa.gov/docs/suzaku/analysis/abc/}}.

For XIS, before extracting spectral products, we used the correction tools{\footnote{http://www-x.phys.se.tmu.ac.jp/~syamada/ana/suzaku}} described in \citet{yam2012} to correct the mean position shift of the source, and to estimate the level of pileup for XIS0 and XIS3. New attitude files were generated and were used as input for XISCOORD to create new cleaned event files. Unlike the pileup estimation in \citet{rei2012} in which they reported a maximum pileup fraction of 2\% using the script PILE\_EST\footnote{https://space.mit.edu/ASC/software/suzaku/pest.html} \citep{dav2001}, we evaluated that the pileup fraction of the source is larger than 3\% within the circle with a radius of 22.8 pixel (1\% at 58.7 pixel) from the source center for XIS0 and larger than 3\% within the circle with a radius of 26.0 pixel (1\% at 60.6 pixel) for XIS3, respectively. 
The Pileup Correction Tool also created relevant region files. \citet{yam2012} recommended using the X-ray events from the region outside the radius with pileup fraction of 3\% or 1\% for spectral analysis. Here we extracted spectra from an annular region which excluded events with pileup fraction higher than 3\%.

The background spectra were extracted from a circular region with a radius of 100$''$ which is away from the source but still on the same chip. Then, we used XISRMFGEN and XISARFGEN to generate the new ancillary response files and redistribution matrix files for both XIS0 and XIS3, respectively. Finally, in order to increase the signal-to-noise ratio (S/N), we combined their spectra, backgrounds and response files. We added systematic errors of 1\% to the spectrum to take into account the calibration uncertainties.  

For HXD, the spectrum (12-70 keV) observed by PIN diodes is used, while the spectrum (40-600 keV) detected by GSO scintillators is abandoned due to the low S/N. We obtained the appropriate response file (ae\_hxd\_pinxinome11\_20110601.rsp) and the non X-ray background (NXB) file from the HXD team \citep{fuk2009}. We extracted the background spectrum based on the NXB file and the total spectrum based on the cleaned events from the same good-time intervals. The 7\% dead time of the observed spectrum was corrected. The contribution from the cosmic X-ray background (CXB), which contributes 5\% of the PIN background, was estimated by simulation with ``fakeit'' command in XSPEC, in which the model \citet{bol1987} and the ``flat'' response (ae\_hxd\_pinflate11\_20110601.rsp ) were used. Furthermore, the normalization of the model was adjusted so that the contribution from CXB is 5\%. Then, the NXB and CXB were added together to obtain the total background spectrum.

For XIS data, due to the calibration issues below $\sim$1.0 keV, we only use the data between the 1.2-10.0 keV energy range. The 1.6-2.0 keV and 2.2-2.4 keV energy ranges are also excluded due to the Si K and the Au M edge at $\sim$1.8 and $\sim$2.2 keV, respectively. For HXD PIN data, we restrict its energy range between 15.0-50.0 keV. When both spectra are simultaneously fitted, a normalization factor of 1.16 is adopted as advised in the \emph{Suzaku} data analysis guide.

\subsection{\emph{RXTE} observations}
\label{subsec:rxte}
The standard \emph{RXTE} products (the source, background and response files) which were reduced from observations MJD 55818.84 and MJD 55819.16, were downloaded from the HEASARC data archive. The PCA spectra are used, while the HEXTE data are discarded due to the low S/N. The PCA were extracted from PCU2, the best-calibrated detector. As a tradition, we added 0.6\% systematic uncertainties and rebinned the spectra with at least 25 photons within each bin. Following the spectral analysis of PCA spectra, we restrict our analysis in the energy range of 3.0-25.0 keV \citep{mil2009}.

\section{Spectral Analysis and Results}
\label{sec:results} 
All spectra were analysed using XSPEC version 12.9.0g \citep{arn1996}. In order to model the galactic absorption, we used the {\tt TBabs} \citep{wil2000} model. The solar abundances from \citet{wil2000} and the photoelectric cross-sections from \citet{ver1996} were adopted. We fixed the column density to $0.2 \times 10^{22}$ cm$^{-2}$ \citep{ken2011} which was given by fitting the \emph{Swift}/XRT observation. All uncertainties calculated for specified parameters in this paper are at 90\% confidence level ($\Delta \chi ^2=2.71$), unless noted particularly. 

\subsection{Preliminary \emph{Suzaku} Spectral Analysis}
\label{subsec:pre}
\begin{figure}
\centering
	\includegraphics[scale=0.50,angle=0]{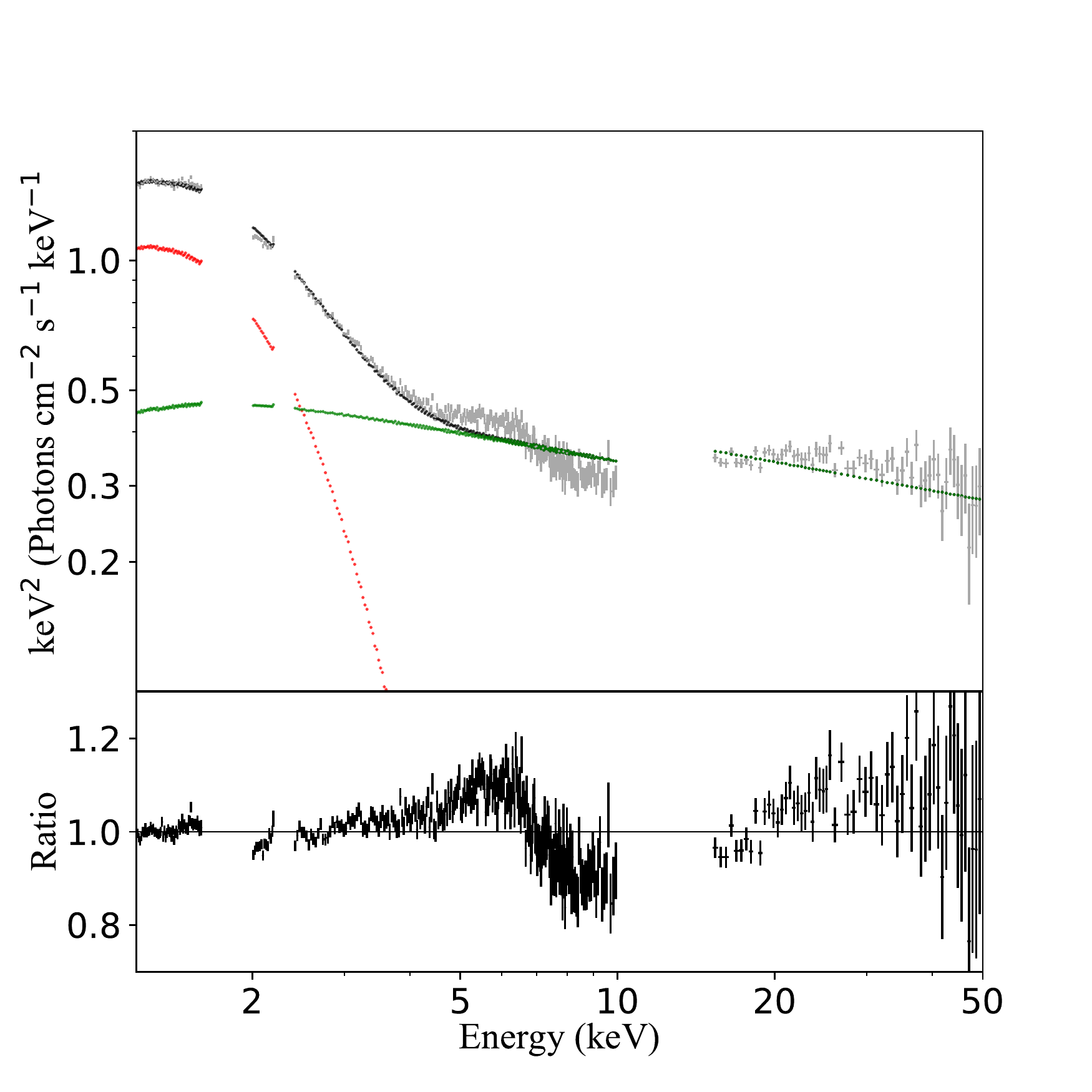}
    \caption{Unfolded \emph{Suzaku} spectra fitted with ignoring the 4-7 keV energy band, but including when plot. In the top panel, the total, {\tt diskbb} and {\tt powerlaw} components are black, red and green dotted lines, respectively. In the bottom panel, the curvature in the data-to-model ratio plot shows the clear signature of the disc reflection.}
    \label{fig:pl}
\end{figure}

We performed preliminary spectral fits to \emph{Suzaku} spectrum with a model consisting of a power-law component ({\tt powerlaw}) and a multi-temperature blackbody ({\tt diskbb}, \citealt{mit1984}), in specific, {\tt TBabs(diskbb+powerlaw)}. The 4-7 keV energy band were excluded to avoid the contribution from the potential broad Fe K$\alpha$ line. 
The best-fitting model as well as the ratio of the model to data are shown in Figure {\ref{fig:pl}}. It is clear from Figure {{\ref{fig:pl}}} that the model does not fit the data well with $\chi^2_{\nu}$ = 1.444 (2924.13/2025). A skewed broad iron line profile, iron absorption edge and a Compton hump are shown in the residues which could be the signature of the disc reflection.

The best-fitting photon index of {\tt powerlaw} is 2.22 $\pm$ 0.01 and the temperature of {\tt diskbb} is 0.439 $\pm$ 0.002 keV. The total 2.0-20.0 keV unabsorbed flux is $\sim1.8 \times 10^{-9}$ erg cm$^{-2}$ s$^{-1}$, 22\% of which is attributed to the thermal emission. \citet{rei2012} reported a photon index of $\sim$2, which is slightly smaller than our result. The fraction of the thermal emission in the total unabsorbed flux ($\sim1.5 \times 10^{-9}$ erg cm$^{-2}$ s$^{-1}$) estimated in their paper is 26\%. Our steeper power-law and higher flux maybe attributed to the pileup correction process. Given a distance of 10 kpc and a mass of 8 $M_{\odot}$ \citep{rus2014b}, we calculated the Eddington ratio $L/L_{\rm{Edd}}$ to be 11\%, suggesting a geometrically thin and optically thick accretion disc \citep{mcc2006}. Thus, it is reasonable to assume that the inner radius of the disc is at the innermost stable circular orbit ($R_{\inner}$ = $R_{\ISCO}$).

We also tried two other models, {\tt cutoffpl} and {\tt nthComp} \citep{zdz1996,zyc1999}, to fit the data. The model {\tt cutoffpl} is a phenomenological power-law model with e-fold at high energies. The model {\tt nthComp} is a physically motivated thermal Comptonization model in which the thermal seed photons from the disc gain energies by interacting with electrons in the high-temperature corona. The {\tt laor} \citep{lao1991} and {\tt smedge} \citep{ebi1994} models were added to account for the reflection feature found in the ratio plot (Figure \ref{fig:pl}). These two models are {\tt TBabs*smedge(diskbb+cutoffpl+laor)} (hereafter M$_{\rm{off}}$) and {\tt TBabs*smedge(diskbb+nthComp+laor)} (hereafter M$_{\rm{nth}}$), respectively. The central energy of the broad iron line was constrained between 6.40 and 6.97 keV for the {\tt laor} model, including all the possible ionization states of iron. The edge could change from 7.0 to 9.0 keV and the smearing width was fixed at 7 keV for the {\tt smedge} model.

The M$_{\rm{off}}$ and M$_{\rm{nth}}$ can fit the data equally well with reduced chi-square of 1.062 and 1.065 for the same degrees of freedom, respectively. The best-fitting thermal temperature, normalization and photon index of the two models are in agreement with each other, with $kT_{\rm{disc}}$ = 0.433 $\pm$ 0.002 keV, $N_{\rm{disc}}$ = 5842 $_{-77}^{+96}$ and $\Gamma$ = 2.156 $\pm$ 0.008 in M$_{\rm{off}}$, and $kT_{\rm{disc}}$ = 0.433 $\pm$ 0.002 keV, $N_{\rm{disc}}$ = 5842 $_{-104}^{+109}$, and $\Gamma$ = 2.150 $\pm$ 0.016 in M$_{\rm{nth}}$. In both cases, the high-energy cutoff parameters are not constrained and only upper limits (300 keV) can be given. An inner radius smaller than $\sim$2.9 $R_{\rm{g}}$ in the {\tt laor} model indicates that the reflection emission arises from the innermost region around a rapidly rotating black hole. Moreover, the normalization of the {\tt diskbb} model could also provide a measurement of the inner radius of the accretion disc ($R_{\rm{in}}$ = $D_{10 {\rm{kpc}}}[N_{\rm{disc}}/{\rm{cos}}(i)]^{1/2}$, where $D_{10{\rm{kpc}}}$ is the source distance in unit of 10 kpc). With the inclination angle of 4-15 degrees  \citep{rus2014a} and the source distance of 4-10 kpc \citep{rus2014b}, our fits ($N_{\rm{disc}}$ = 5842 $_{-104}^{+109}$) indicate that the inner disc extends to 2.58 - 6.44 $R_{\rm{g}}$, which is in agreement with the value obtained by {\tt laor} model.

\subsection{Relativistic reflection Models for \emph{Suzaku} and \emph{RXTE} Spectra}
\label{subsec:phy}

\begin{table*}
    \caption{Best-fitting Parameters with Relativistic Models}
    \begin{threeparttable}[b]
    \begin{center}
    \label{tab:rel}
    \footnotesize
        \begin{tabular}{ccccc}
        \toprule
        Parameter & Model 1  & Model 2 & Model 3 & Model 4\\
        \midrule
        \multicolumn{5}{c}{\emph{Suzaku}}\\
        \midrule
        \multicolumn{5}{l}{Multi-temperature blackbody}\\
        \specialrule{0em}{1pt}{1.3pt}
        
        $kT_{\rm{disc}}$ (keV)	&0.436 $_{-0.002}^{+0.003}$	&0.4380 (0.0004)$^a$	&0.434 (0.003)	&0.412 $_{-0.002}^{+0.010}$	\\
        \specialrule{0em}{1pt}{1.3pt}					
        $N_{\rm{disc}}$	&5660 $_{-168}^{+129}$	&5588 (18)	&5887 $_{-191}^{+164}$	&6484 $_{-715}^{+477}$	\\
        \midrule					
        \multicolumn{5}{l}{Power-law continuum plus relativistic reflection}\\					
        \specialrule{0em}{1pt}{1.3pt}					
        $\Gamma$	&2.11 $_{-0.01}^{+0.02}$	&2.109 (0.002)	&2.00 $_{-0.01}^{+0.06}$	&2.02 $_{-0.02}^{+0.01}$	\\
        \specialrule{0em}{1pt}{1.3pt}					
        $h$ $(R_{\rm{g}})$	&...	&2.36 $_{-0.17}^{+0.53}$	&...	&...	\\
        \specialrule{0em}{1pt}{1.3pt}					
        $q_{\rm{in}}$	&6.42 $_{-1.83}^{+1.10}$	&...	&4.38 $_{-0.59}^{+0.57}$	&> 5.95 	\\
        \specialrule{0em}{1pt}{1.3pt}					
        $R_{\rm{br}}$ $(R_{\rm{g}})$	&4.45 $_{-0.64}^{+0.72}$	&...	&5.34 $_{-0.76}^{+0.72}$	&4.01 $_{-0.29}^{+0.36}$	\\
        \specialrule{0em}{1pt}{1.3pt}					
        $a_*$	&0.88 $_{-0.04}^{+0.03}$	&0.934 (0.005)	&0.91 $_{-0.04}^{+0.03}$	&0.88 $_{-0.03}^{+0.02}$	\\
        \specialrule{0em}{1pt}{1.3pt}					
        $i$ (deg)$^b$	& 5 $^{+4}$	& 15 $_{-1}$	& 4 $^{+3}$	& 5 $^{+3}$	\\
        \specialrule{0em}{1pt}{1.3pt}					
        $A_{\rm{Fe}}$	&4.99 $_{-0.68}^{+1.02}$	& > 9.35	& > 6.62	&1(f)$^{c}$	\\
        \specialrule{0em}{1pt}{1.3pt}					
        $\log\xi_1$	&3.67 $_{-0.12}^{+0.05}$	&3.63 (0.05)	&4.23 $_{-0.13}^{+0.12}$	&2.50 $_{-0.09}^{+0.11}$	\\
        \specialrule{0em}{1pt}{1.3pt}					
        $R_{\rm{ref}}$	&0.45 $_{-0.06}^{+0.12}$	&1.33 (0.02)	&0.65 $_{-0.25}^{+0.45}$	&...	\\
        \specialrule{0em}{1pt}{1.3pt}					
        $\log\xi_2$	&...	&...	&3.47 $_{-0.06}^{+0.10}$	&...	\\
        \specialrule{0em}{1pt}{1.3pt}					
        $\log n_{e}$	&...	&...	&...	& > 21.83	\\
        \specialrule{0em}{1pt}{1.3pt}					
        $N_{\rm{\tt cutoffpl}}$	&...	&...	&...	&0.25 (0.01)	\\
        \specialrule{0em}{1pt}{1.3pt}					
        $N_{\rm{\tt relxill_1}}$ $(\times10^{-2})$	&0.38 $_{-0.04}^{+0.03}$	&6.24 $_{-0.01}^{+1.96}$	&0.21 $_{-0.06}^{+0.08}$	&...	\\
        \specialrule{0em}{1pt}{1.3pt}					
        $N_{\rm{\tt relxill_2}}$ $(\times10^{-2})$	&...	&...	&0.119 $_{-0.02}^{+0.08}$	&...	\\
        \specialrule{0em}{1pt}{1.3pt}					
        $N_{\rm{\tt reflionx\_hd}}$	&...	&...	&...	&1.17 $_{-0.10}^{+0.38}$	\\
        \specialrule{0em}{1pt}{1.3pt}					
        \midrule					
        \multicolumn{5}{c}{\emph{RXTE}}\\					
        \midrule					
        Cross-Normalization constant (relative to \emph{Suzaku}/XIS)\\					
        C	&1.035 $_{-0.005}^{+0.004}$	&1.035 (0.004)	&1.034 (0.005)	&1.035 (0.005)	\\
        \midrule					
        $\chi^2/\nu$	& 2118.76/2115	& 2144.58/2116	& 2107.01/2113	& 2093.68/2115	\\
        $\chi^2_{\nu}$	& 1.002	& 1.014	& 0.997	& 0.990	\\
        \bottomrule
        \end{tabular}
     \begin{tablenotes}
        \item \textbf{Notes.} The best-fitting parameters obtained by modelling \emph{Suzaku} and \emph{RXTE} observations. Model 1 is {\tt TBabs*(disbkk+relxill)} assuming an extended corona with broken power-law emissivity profile. Model 2 is {\tt TBabs*(disbkk+relxilllp)} assuming a lamp-post corona. Model 3 is {\tt TBabs*(disbkk+relxill$_1$+relxill$_2$)}, namely we add another relativistic reflection component in Model 1. Model 4 is {\tt TBabs(diskbb+cutoffpl+relconv*reflionx$\_$hd)}, in which the electron density is a free parameter. For models {\tt relxill(lp)} or {\tt relconv} in Model 1, 2 and 4, the inner radius $R_{\rm{in}}$ is equal to $R_{\rm{ISCO}}$, and the outer radius $R_{\rm{out}}$ is fixed at its default value: $R_{\rm{out}}$ = 400 $R_{\rm{g}}$. In Model 3, in the first reflection (R1), $R_{\rm{in}}$ is equal to $R_{\rm{ISCO}}$ and $R_{\rm{out}}$ is equal to $R_{\rm{br}}$; in the second (R2), $R_{\rm{in}}$ is equal to $R_{\rm{out}}$ in R1, $R_{\rm{out}}$ is equal to 400 $R_{\rm{g}}$, parameters $q_{\rm{in}}$, $q_{\rm{out}}$, and $R_{\rm{br}}$ are fixed at their default values, i.e. $q_{\rm{in}}$ = 3, $q_{\rm{out}}$ = 3, and $R_{\rm{br}}$ = 15 $R_{\rm{g}}$, and $\log\xi_2$ and $N_{\rm{\tt relxill_2}}$ are free. The remaining common parameters are linked together for R1 and R2.
        
        \item[a] One digit enclosed in parentheses implies that the up error equal to the low.
        \item[b] The inclination angle pegs at its lower or upper limit (4-15 degrees) we set.
        \item[c] The value followed by f in parentheses implied that it is fixed at some value.
    \end{tablenotes}
    \end{center}
    \end{threeparttable}
\end{table*}
\begin{figure*}
\centering
	\includegraphics[scale=0.50,angle=0]{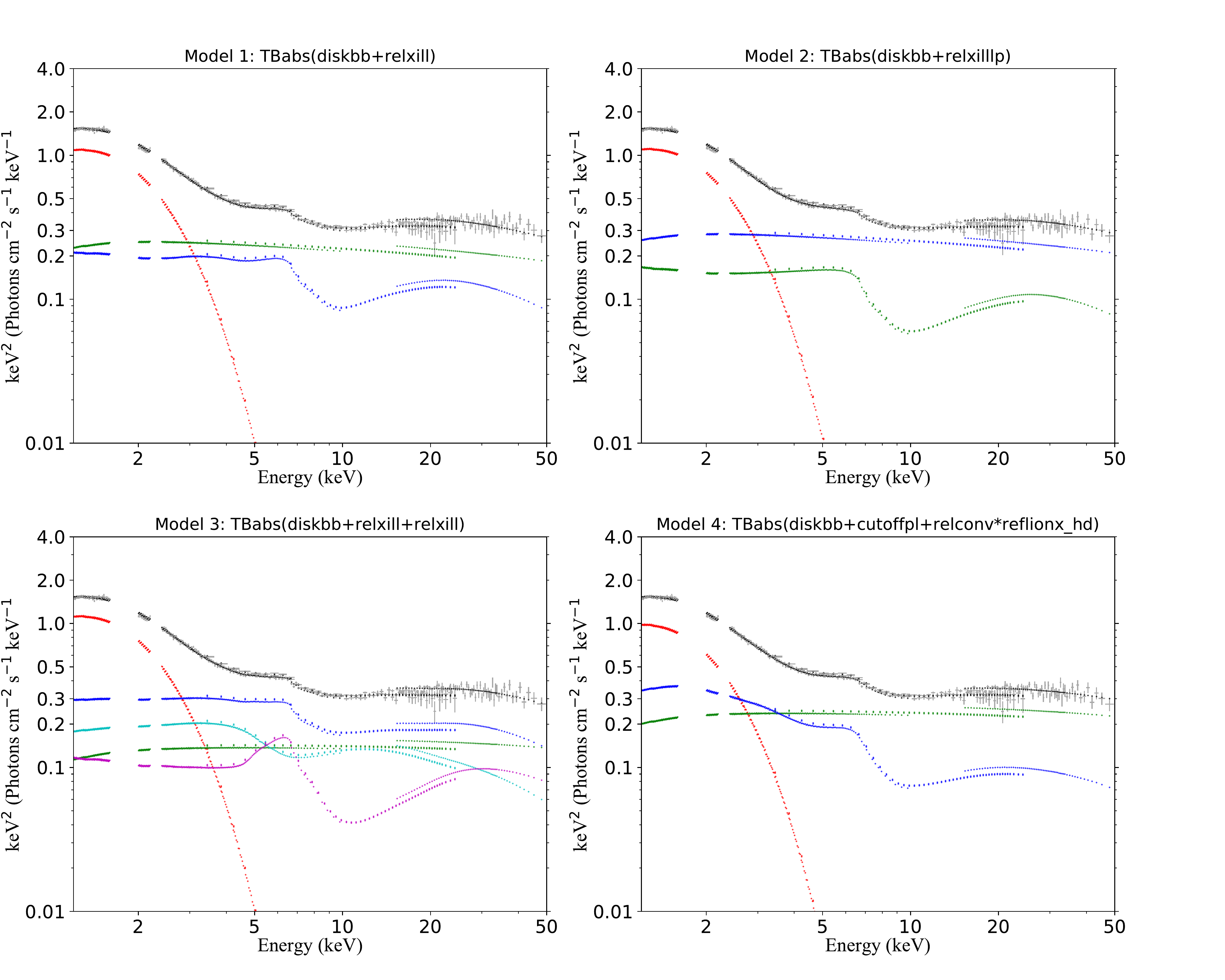}
    \caption{Unfolded \emph{Suzaku} and \emph{RXTE} spectra fitted with four models. In all panels, the total, the thermal, the power-law and the reflection components are black, red, green and blue dotted lines, respectively. In the bottom left panel, we use cyan and purple dotted lines to describe the relativistic reflection component in the inner disc and the outer disc, respectively, and the blue to describe the sum of the two reflection components.}
    \label{fig:comp}
\end{figure*}

\begin{figure*}
\centering
	\includegraphics[scale=0.50,angle=0]{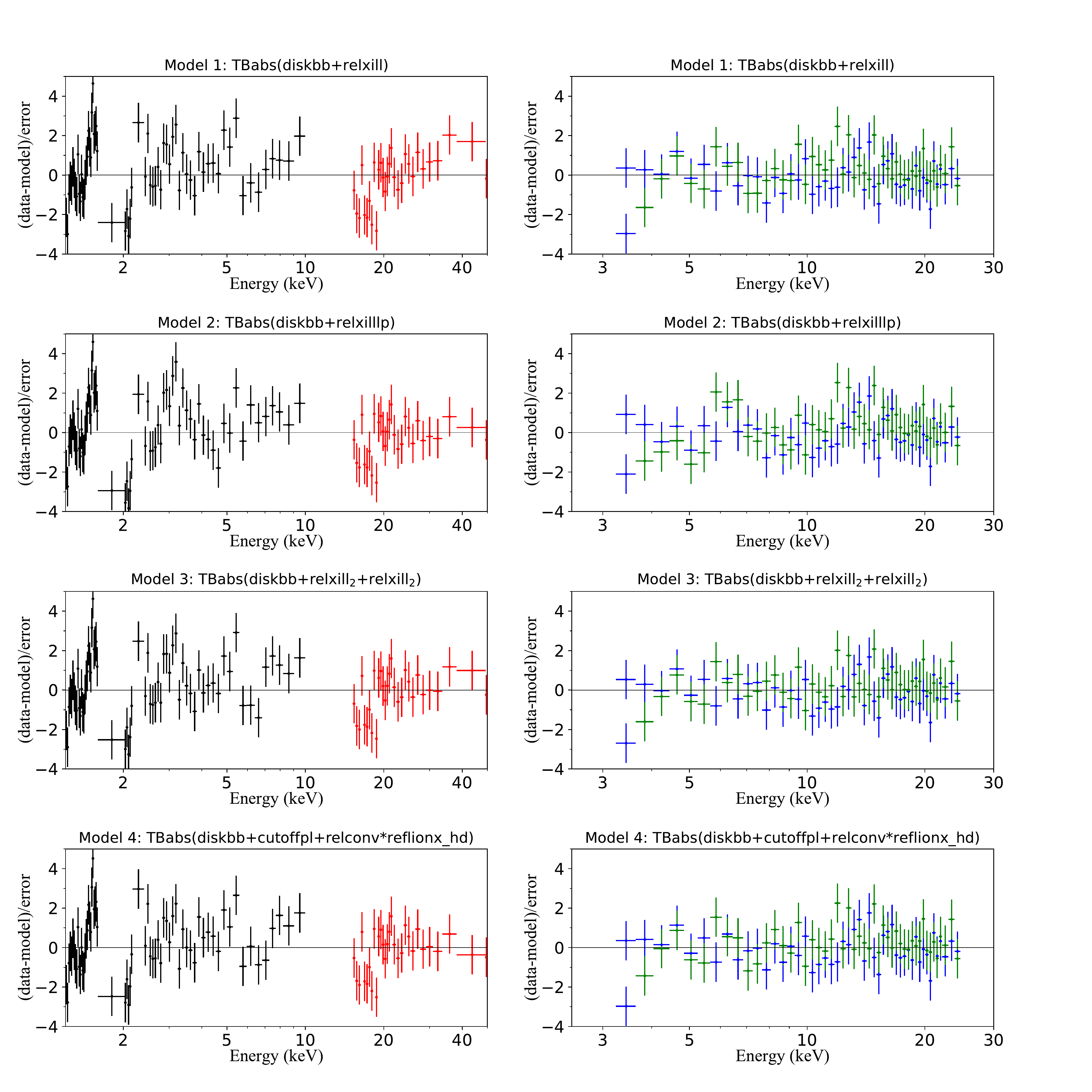}
    \caption{The residuals with 1$\sigma$ for four models are shown in plot. The black and red represent the result of fit to \emph{Suzaku}/XIS and PIN spectra, respectively (left panels). The green and blue represent the result of fit to \emph{RXTE}/PCA spectra obtained in MJD 55818.84 and 55819.16, respectively (right panels). These spectra were fitted together, but are shown separately for clarity. The XIS and HXD spectra are binned in XSPEC only for visual clarity.}
    \label{fig:delchi}
\end{figure*}

In order to measure the spin of the black hole in MAXI J1836-194, we replaced the {\tt laor}, {\tt smedge}, and power-law models with the {\tt relxill} model to fit the reflection emission and the power-law continuum. In addition, to better constrain the reflection component, we extended the energy band to 15 keV by fitting the \emph{Suzaku} and \emph{RXTE} simultaneously.

A multiplicative constant is included to account for the differences in the flux calibration between the \emph{RXTE}/PCA and the \emph{Suzaku}/XIS.
The inclination angle $i$ was bounded between 4-15 degrees based on the optical measurements \citep{rus2014a}. The power-law continuum is described with the exponential cut-off power-law. The inner radius $R_{\rm{in}}$ is equal to $R_{\rm{ISCO}}$. The outer radius $R_{\rm{out}}$ is fixed at its default value: $R_{\rm{out}}$ = 400 $R_{\rm{g}}$. We noted, that as demonstrated in Section {\ref{subsec:pre}}, changing it to a thermal comptonization model will not affect our results. 
The high-energy cutoff $E_{\rm{cut}}$ is equal to $\sim(2-3)$ $kT_{e}$, in which $kT_{e}$ is the electron temperature of the corona. $E_{\rm{cut}}$ is fixed at its default value of 300 keV.

The power-law continuum was assumed to come from an extended corona (Model 1). A broken power-law emissivity profile was adopted. The outer index was fixed at its canonical value ($q_{\rm{out}}=3$), while the inner index $q_{\rm{in}}$ and the break radius $R_{\rm{br}}$ were set free. This model gives a much better fit with $\chi^2_{\nu}$ = 1.002 (2118.76/2115), compared to the initial fit with {\tt TBabs(diskbb+powerlaw)} in Section \ref{subsec:pre}. The best-fitting results for each parameter can be found in Table \ref{tab:rel}. The best-fitting model is shown in the top left panel in Figure {\ref{fig:comp}}. The residuals with 1$\sigma$ of the best-fitting are shown in the top two panels in Figure {\ref{fig:delchi}}.

The temperature and the normalization of the thermal emission are 0.436 $_{-0.002}^{+0.003}$ keV and 5660 $_{-168}^{+129}$, respectively. The photon index is 2.11 $_{-0.01}^{+0.02}$. The inner index of the emissivity profile is 6.42 $_{-1.83}^{+1.10}$ and the break radius is 4.45 $_{-0.64}^{+0.72}$ $R_{\rm{g}}$, implying that the corona is compact and the flux of the reflected emission decreases dramatically within the break radius. The best-fitting spin parameter is at a moderate value of 0.88 $_{-0.04}^{+0.03}$. The fit gives an upper limit of 9 degrees for the inclination angle. The iron abundance $A_{\rm{Fe}}$ is 4.99 $_{-0.68}^{+1.02}$. The logarithmic ionization state is 3.67 $_{-0.12}^{+0.05}$. The reflection fraction which defines the photon fraction hitting the accretion disc \citep{dau2014} is 0.45 $_{-0.06}^{+0.12}$. When the outer index is free, it is constrained to be 3.31 $_{-0.04}^{+0.05}$, and the inner index and the iron abundance only obtained their lower limit of 7.46 and 7.29, respectively. The values of the inner index and the iron abundance are too large to be considered as physical. In any event, the value of the spin parameter we cared most is not affected. Therefore, we will set the outer index at 3 in the rest fits for simplicity.

In previous studies, a narrow Fe K$\alpha$ line, which could be produced in the region far away from the central black hole, was already detected in the X-ray spectra of some black hole X-ray binaries, such as GX 339-4 \citep{gar2015} and Cyg X-1 \citep{tom2018}. However, there is no clear evidence for such a narrow line in the X-ray spectrum of MAXI J1836-194 (see top two panels of Figure \ref{fig:delchi}).
To further test the significance of this component, we added the {\tt xillver} model, which is used to model unblurred reflection component, into Model 1. We found that the inclusion of this component did not significantly improve the fit with $\chi^2_{\nu}$ = 1.002 (2118.39/2114). The intensity of the relativistic reflection is also 30 times stronger than it. Additionally, it did not affect the values of other parameters, such as the spin, the inclination angle, and the iron abundance. Therefore, we did not include this distant reflection in our fits.

As to the geometry of the corona, it is still unclear, but there are two popular models, extended geometry \citep{wil2012b} and lamp-post configuration \citep{mat1991, mar1996}. In Model 1, we have assumed a broken power-law emissivity to explore the extended corona, in which case we found that a compact corona is required. In order to test the effect of different geometry on the spin, we also tried to fit the data with a lamp-post configuration using the model {\tt relxilllp} (Model 2). The lamp-post model leads to a slightly worse fit with $\chi^2_{\nu}$ = 1.014 (2144.58/2116) compared to Model 1. The best-fitting results of Model 2 are shown in Table \ref{tab:rel}. The model components are shown in the top right panel in Figure \ref{fig:comp} and the residuals with 1$\sigma$ are shown in the second top two panels in Figure \ref{fig:delchi}. 

The fitted value for the parameters in Model 2 are listed here. The temperature and the normalization of the thermal disc are 0.4380 $\pm$ 0.0004 keV and 5588 $\pm$ 18, respectively. The photon index is 2.109 $\pm$ 0.002. The logarithmic ionization state is 3.63 $\pm$ 0.05. These best-fitting parameters are consistent with the results found in Model 1. The height of the point source is 2.36 $_{-0.17}^{+0.53}$ $R_{\rm{g}}$, which indicates that a compact corona is located closely to the black hole. The spin parameter of the black hole is 0.934 $\pm$ 0.005, slightly higher than that in Model 1. The inclination angle is larger than 14 degrees but pegged at the upper limit of 15 degrees. However, the inclination angle was constrained to be {$\sim$}16-19 degrees when the upper bound is set at a much larger value, which is still consistent with a low disc inclination angle. The iron abundance $A_{\rm{Fe}}$ is larger than 9.35. Moreover, Model 2 requires a stronger illumination of $N_{\rm{relxill}}$ = (6.24 $_{-0.01}^{+1.96}$) $\times 10^{-2}$ and reflection fraction of $R_{\rm{ref}}$ = 1.33 $\pm$ 0.02 compared to those in Model 1.
 
It is suggested that the constant ionization in current models may lead to a steep emissivity index. To explore the effect of constant ionization on steep emissivity index, we fit the data with a dual-relativistic reflection model, i.e., a second relativistic reflection component was added in Model 1. This new model is defined as Model 3. We use R1 and R2 to differentiate the two {\tt relxill} models. In R1, the outer radius $R_{\rm{out}}$ is equal to the break radius $R_{\rm{br}}$, while in R2 the inner radius $R_{\rm{in}}$ is linked to $R_{\rm{out}}$ in R1. The other parameters, $q_{\rm{in}}$, $q_{\rm{out}}$, and $R_{\rm{br}}$ in R2, are fixed at their default values, i.e. $q_{\rm{in}}$ = 3, $q_{\rm{out}}$ = 3, and $R_{\rm{br}}$ = 15 $R_{\rm{g}}$. The ionization parameters in R1 and R2 are free. The normalization parameter in R2 is also free. The remaining parameters in R1 and R2 are linked together. 

Compared with Model 1, Model 3 provides a better statistics with $\chi^2_{\nu}$ = 0.997 (2107.01/2113). The best-fitting parameters of Model 3 are presented in Table \ref{tab:rel}. The model components are shown in the bottom left panel in Figure \ref{fig:comp} and the residuals with 1$\sigma$ of the best-fitting are shown in the third top two panels in Figure \ref{fig:delchi}.
For Model 3, the temperature and the normalization of the thermal disc are 0.434 $\pm$ 0.003 keV and 5887 $_{-191}^{+164}$, respectively. The photon index is 2.00 $_{-0.01}^{+0.06}$, respectively. The spin parameter of the black hole is 0.91 $_{-0.04}^{+0.03}$. The inclination angle is smaller than 7 degrees. The iron abundance $A_{\rm{Fe}}$ is larger than 6.62. These parameters are consistent with the best-fitting parameters in Model 1. The best-fitting emissivity index of R1 is 4.38 $_{-0.59}^{+0.57}$ with the break radius of 5.34 $_{-0.76}^{+0.72}$ $R_g$, which indicate a more extended corona than that shown in Model 1. The logarithmic ionization state of R1 and R2 are 4.23 $_{-0.13}^{+0.12}$ and 3.47 $_{-0.06}^{+0.10}$, respectively, which may indicate ionization gradient on the surface of the disc. 

We noticed that the iron abundances in all three models prefer supersolar. Forcing $A_{\rm{Fe}}$ at the solar abundance in Model 1 will lead to a worse fit ($\Delta\chi^{2}\sim$156.63 with 1 degree of freedom), but the spin parameter $a_*$ is 0.79 $_{-0.10}^{+0.06}$, which still indicates a moderate rotating black hole. We also tried a new version of {\tt reflionx} model ({\tt reflionx\_{hd}}) with the electron density as a free parameter \citep{tom2018}. The iron abundance is fixed at solar abundance. The {\tt relconv} is used to account for the relativistic effect on the reflection emission, while the {\tt cutoffpl} is adopted to model the power-law emission. This model is defined as Model 4. Model 4 only improve the fit slightly with $\chi^2_{\nu}$ = 0.990 (2093.68/2115). The best-fitting parameters are presented in Table \ref{tab:rel}. The model components are shown in the bottom right panel in Figure \ref{fig:comp} and the residuals with 1$\sigma$ of the best-fitting are shown in the bottom two panels in Figure \ref{fig:delchi}. It does not show any significant residual either in \emph{Suzaku} or in \emph{RXTE} data. The spin parameter of the black hole is 0.88 $_{-0.03}^{+0.02}$. The best-fitting inclination angle is smaller than 8 degrees. The logarithmic ionization parameter is 2.50 $_{-0.09}^{+0.11}$, which is smaller than that found in other models. The most important thing is that the density is larger than 10$^{21}$ cm$^{-3}$, which is significantly larger than the assumed disc density of 10$^{15}$ cm$^{-3}$ in models {\tt relxill} and {\tt relxilllp}.

\section{Discussion}
\label{sec:dis}
In this paper, we analysed the broad-band X-ray data of MAXI J1836-194 in the intermediate state, using the simultaneously observed X-ray data from \emph{Suzaku} and \emph{RXTE}.
The unabsorbed Eddington ratio is approximately 11\% in which the accretion disc is geometrically thin and optically thick, suggesting that the inner edge of the disc has already reached the ISCO radius. The sophisticated relativistic reflection models ({\tt relxill(lp)} and {\tt reflionx\_hd}) are used to fit the data and measure the spin of the black hole. 

\begin{figure*}
\centering
	\includegraphics[scale=0.50,angle=0]{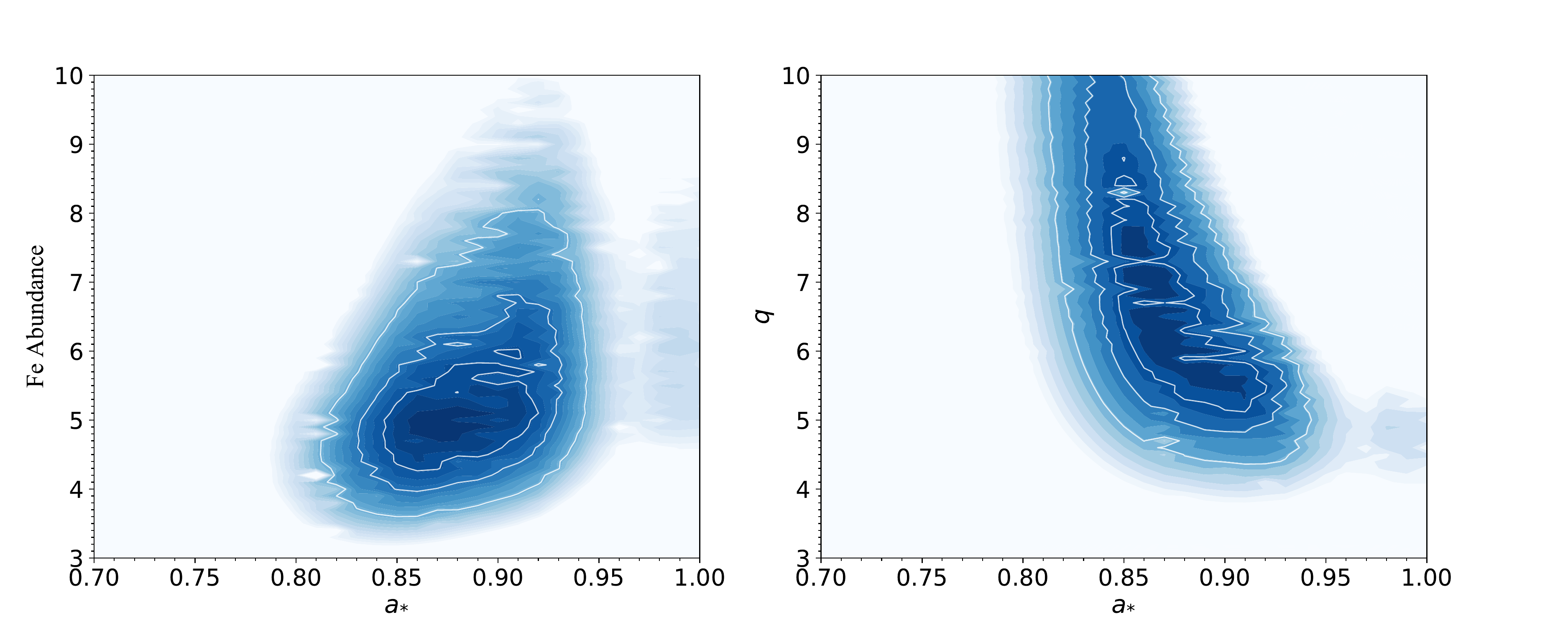}
    \caption{The correlation between the spin parameter $a_*$ and the iron abundance $A_{\rm{Fe}}$ and the correlation between the spin parameter $a_*$ and the inner index $q_{\rm{in}}$ of emissivity are calculated with the command "steppar" based on Model 1. For visual guide, probability contours of 68\%, 90\%, and 99\% are shown in white solid lines. }
    \label{fig:aFeq}
\end{figure*}

We fitted the data with four models. In Model 1, we assumed an extended corona covering above the accretion disc with a broken power-law emissivity. In Model 2, we assumed a lamp-post configuration in which a point-like corona is located above the black hole. In Model 3, to explore the effect of ionization gradient, we fitted the data with a model consisting of two relativistic reflection components. One of the two components is used to fit the reflection from the inner region of the accretion disc, while the other one is used to account for the radiation from the outer region. In Model 4, we applied a new version of {\tt reflionx} with the electron density to be a free parameter. As can be seen from Figure \ref{fig:delchi}, all the four models can fit the data well, giving acceptable fits.

According to best-fitting results shown in Table \ref{tab:rel}, a thermal component with the temperature of approximately 0.4 keV is required in all models. But the thermal emission below 5 keV in the Model 4 is slightly smaller than that in other three models (see Figure \ref{fig:comp}). The fact that the index of the {\tt powerlaw} component is $\sim$2, together with the disc fraction of 26\%, implies that the source is in the intermediate state. However, the power-law index is slightly flatter in Model 3-4. For Model 3, this may be due to the more contribution from thermal emission, while for Model 4, it is led by the more contribution  from the reflection (see Figure \ref{fig:comp}) at soft X-ray band. The inclination angle of the disc is very low, which is consistent with the optical measurement made by VLT \citep{rus2014a}.

\subsection{Steep inner emissivity index}
\label{subsec:q}
A steep inner emissivity index (6.42 $_{-1.83}^{+1.10}$) and a small break radius (4.45 $_{-0.64}^{+0.72}$ $R_{\rm{g}}$), which is common in X-ray binary and AGN, is found in Model 1 (assuming the broken power-law emissivity). We showed the contour plot for the spin and the emissivity index for Model 1 in the right panel in Figure \ref{fig:aFeq}. A mildly inverse relationship between these two parameters is shown. Steep emissivity indexes ($q_{\rm{in}}$ > 4) are required for any value in the narrow spin range. Fixing $q_{\rm{in}}$ at 3 and $R_{\rm{br}}$ at 15 $R_{\rm{g}}$ will lead to a worse fit with $\chi^2_{\nu}$ = 1.033 (2186.91/2117). The spin parameter $a_*$ is inferred to be larger than 0.93. 

The steep inner index may be explained with a compact corona close to the black hole. Our Model 2 (assuming a point-like corona) is used to test this explanation. As what we expected, a very low height ($h$ = 2.36 $_{-0.17}^{+0.53}$ $R_{\rm{g}}$) of the compact corona is found. The light-bending effect focuses more photons on the inner region of the disc, which steepens the inner index. This effect may also lead the reflection fraction parameter larger than unity. In our case, the reflection fraction $R_{\rm{ref}}$ is 1.33 $\pm$ 0.02. However, we found that the reflection component is smaller than that in Model 1 (compare the top two panels in Figure {\ref{fig:comp}}), and the reflection component is weaker than the power-law component. We think it is because, in this configuration, the spin parameter $a_*\sim0.93$ corresponds to the $R_{\rm{ISCO}}$ of 2.10 $R_{\rm{g}}$, then some of the power-law photons that should have hit the accretion disc fall into the gap between the black hole and the disc or directly into the black hole. In addition, we found that the iron absorption edge is larger than that in Model 1 (Figure {\ref{fig:comp}}). The steeper iron edge is due to the suddenly increasing iron abundance ($A_{\rm{Fe}}\sim10$), which is an extremely unphysical value. Moreover, Model 2 is excluded at 99\% confidence level compared to the Model 1. We will then disregard Model 2.

The steep emissivity profile might be caused by the simple assumption of a constant ionization along the radius in the reflection model. The dual-relativistic reflection model (Model 3) found two ionization states ($\log\xi_1$ = 4.23 $_{-0.13}^{+0.12}$ and $\log\xi_2$ = 3.47 $_{-0.06}^{+0.10}$) for different region of the disc. 
Model 3 is favoured over Model 1 as suggested from the F-test (with a p value of 0.003). Model 3 gives an inner emissivity index of 4.38 $_{-0.59}^{+0.57}$ and a break radius of 5.34 $_{-0.76}^{+0.72}$ $R_g$ which indicate a more extended corona than that found in Model 1. Our results give a tentative evidence for ionization gradient along the radius of the disc. According to the bottom left panel in Figure \ref{fig:comp}, we obtain different reflection profiles of the {\tt relxill} model for different ionization parameters. The strength of the two reflection components is comparable. The total reflection is stronger than that in other models. The softer and more ionized reflection component comes from the innermost region. As suggested in \citet{svo2012}, when a constant ionization is assumed in the reflection model, the flux from the innermost region may be significantly underestimated, which then leads to a steep emissivity index. The emissivity found in Model 1 is indeed steeper than that in Model 3, consistent with the prediction from the simulation.

\subsection{High iron abundance}
\label{subsec:AFe}
Model 1 and Model 3 require an extremely supersolar iron abundance. We showed the contour plot for the spin and the iron abundance for Model 1 in the left panel in Figure \ref{fig:aFeq}. The plot suggests a slightly positive correlation between the two parameters. Moreover, The iron abundance is obtained to be 3.5-8 and the spin parameter is well constrained to be 0.81-0.94 at 99\% confidence. \citet{gar2018a} pointed out the most likely explanation for the supersolar iron abundance is the lack of models with very high density. 
The new version of {\tt reflionx} with the electron density free allows the $A_{\rm{Fe}}$ parameter to return back to the solar abundance. Moreover, it provides the best fit among our four models. The high disc density ($n_e$ > 10$^{21}$ ${\rm{cm}}^{-3}$) is suggested by the model. A significantly soft excess bellow 2 keV is found in the bottom right panel in Figure \ref{fig:comp}. \citet{gar2016} demonstrated that the increasing free-free absorption heats the gas, which results in an increase of the gas temperature when density is larger than $10^{17}$ cm$^{-3}$. Then it will increase the flux at low energies (< 2 keV). This effect has been shown in Cyg X-1 \citep{tom2018}. We noticed that the ionization state decreases, which follows its relationship with density (see the formula in Section \ref{sec:Int}). It is noted that the high density does not affect our spin measurement.

\subsection{Spin constraints}
\label{subsec:a}
\begin{figure}
\centering
	\includegraphics[scale=0.50,angle=0]{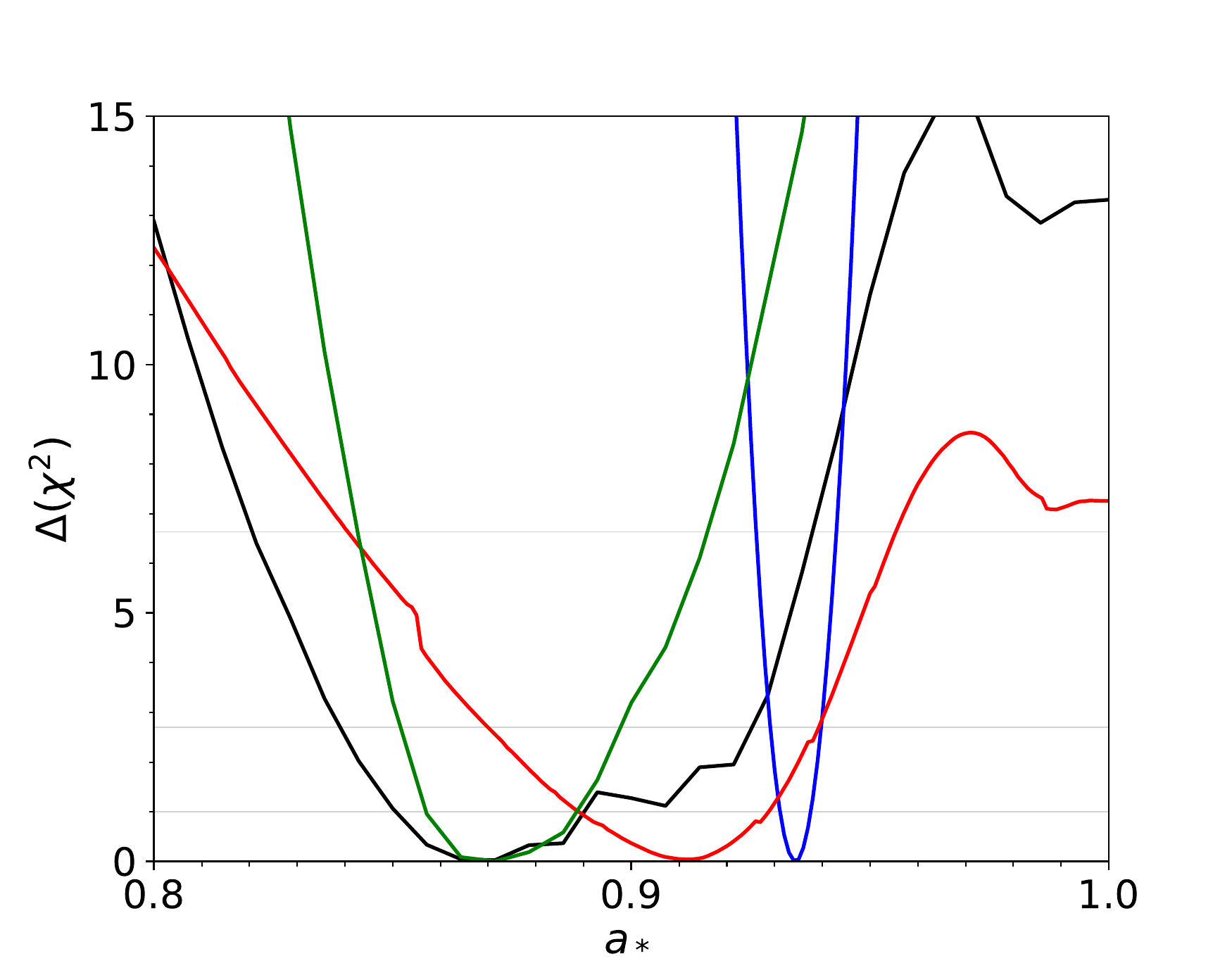}
    \caption{In XSPEC, using "steppar" command to search the best-fit for 20 values of $a_*$ from 0.8 to 1 for Model 1-4. The black, blue, red and green solid line are used to represent the result obtained from Model 1, 2, 3, and 4 successively. The 68\%, 90\% and 99\% confidence intervals are indicated by gray lines.}
    \label{fig:a}
\end{figure}
The spin parameter can be constrained with the best-fitting value $a_*$ = 0.88 $_{-0.04}^{+0.03}$ for Model 1. However, the steep inner emissivity index ($q_{\rm{in}}$ = 6.42 $_{-1.83}^{+1.10}$) and the high iron abundance ($A_{\rm{Fe}}$ = 4.99 $_{-0.68}^{+1.02}$) are also given by Model 1, which calls into question the reliability of the results. When the ionization gradient is considered in Model 3, the inner emissivity index flattens ($q_{\rm{in}}$ = 4.38 $_{-0.59}^{+0.57}$), and the spin parameter changes to 0.91 $_{-0.04}^{+0.03}$. We also used the high-density model to fit the data in Model 4. It is found that the iron abundance can be reduced to the solar abundance, and the spin parameter is $a_*$ = 0.88 $_{-0.03}^{+0.02}$. The spins measured by different models consistently indicate a moderate rotating black hole (see Figure \ref{fig:a}) in MAXI J1836-194. 

Our new spin result is in agreement with that shown in \citet{rei2012}. They used model {\tt refbhb} together with relativistic smearing model to account for the physical processes happening in MAXI J1836-194 system. The {\tt refbhb} model combines the thermal disc and reflection emission in a consistent way, which works well when the disc temperature is so high that the Compton broadening of Fe line is of significance. But in our case, the fraction of the thermal emission to the total unabsorbed flux in the 2.0-20.0 keV band is only 22\%, meaning the thermal emission from the disc is not dominate to the total emission in the observation we work on. The physical model without including the thermal emission we use, therefore, could work well. In the meantime, our model ({\tt relxill}) convolves the reflection with relativistic effect, which is a more advanced method to calculate the emission near the black hole, but {\tt refbhb} needs to be calculated externally.

\section{Conclusion}
\label{sec:con}
 In this paper, to measure the spin of the black hole candidate in binary system MAXI J1836-194 well, we first evaluated and eliminated pile-up effect for \emph{Suzaku} observation, using the newest pipeline provided by \citet{yam2012}, and we also included two simultaneous observations by \emph{RXTE} to increase the spectral energy band. We fit the spectra jointly with the sophisticated relativistic reflection models and a moderate rotating black hole is obtained. The spin parameter is well estimated to be within 0.84-0.94, which agrees well with the spin obtained by \citet{rei2012}. The inclination angle is also constrained, which is consistent with the one obtained by the optical measurement. In the previous fit, there are two potential problems, one is the steep emissivity index, and the other is super-solar abundance. We also tried to solve them. As to the steep emissivity index, it can be solved by introducing an ionization gradient. The supersolar iron abundance can be relieved by increasing the electron density. Of course, the high-density model with solar iron abundance also gives a similar spin with the models which assume the constant density. We hope the future reflection model can incorporate the ionization gradient and allow the higher density to study the spin of black hole.
 
\section*{Acknowledgements}

This research has made use of data obtained from the \emph{Suzaku} satellite, a collaborative mission between the space agencies of Japan (JAXA) and the USA (NASA), and also has made use of standard data products obtained from the \emph{RXTE} satellite and the \emph{RXTE} Guest Observer Facility (GOF). This research has made use of data and/or software provided by the High Energy Astrophysics Science Archive Research Centre (HEASARC), which is a service of the Astrophysics Science Division at NASA/GSFC and the High Energy Astrophysics Division of the Smithsonian Astrophysical Observatory. We thank the useful discussions with Prof. S. Yamada on extracting \emph{Suzaku} spectrum. We thank the high density model provided by Dr. John A. Tomsick. We thank the valuable discussions with Dr. Erlin Qiao. We also thank the reviewer for her valuable comments. Lijun Gou are supported by the National Program on Key Research and Development Project through grant No. 2016YFA0400804, and by the National Natural Science Foundation of China with grant No. Y913041V01, and by the Strategic Priority Research Program of the Chinese Academy of Sciences through grant No. XDB23040100.




\bibliographystyle{mnras}
\bibliography{MAXIJ1836}

\begin{thebibliography}{}
\makeatletter
\relax
\def\mn@urlcharsother{\let\do\@makeother \do\$\do\&\do\#\do\^\do\_\do\%\do\~}
\def\mn@doi{\begingroup\mn@urlcharsother \@ifnextchar [ {\mn@doi@}
  {\mn@doi@[]}}
\def\mn@doi@[#1]#2{\def\@tempa{#1}\ifx\@tempa\@empty \href
  {http://dx.doi.org/#2} {doi:#2}\else \href {http://dx.doi.org/#2} {#1}\fi
  \endgroup}
\def\mn@eprint#1#2{\mn@eprint@#1:#2::\@nil}
\def\mn@eprint@arXiv#1{\href {http://arxiv.org/abs/#1} {{\tt arXiv:#1}}}
\def\mn@eprint@dblp#1{\href {http://dblp.uni-trier.de/rec/bibtex/#1.xml}
  {dblp:#1}}
\def\mn@eprint@#1:#2:#3:#4\@nil{\def\@tempa {#1}\def\@tempb {#2}\def\@tempc
  {#3}\ifx \@tempc \@empty \let \@tempc \@tempb \let \@tempb \@tempa \fi \ifx
  \@tempb \@empty \def\@tempb {arXiv}\fi \@ifundefined
  {mn@eprint@\@tempb}{\@tempb:\@tempc}{\expandafter \expandafter \csname
  mn@eprint@\@tempb\endcsname \expandafter{\@tempc}}}

\bibitem[\protect\citeauthoryear{{Arnaud}}{{Arnaud}}{1996}]{arn1996}
{Arnaud} K.~A.,  1996, in {Jacoby} G.~H.,  {Barnes} J.,  eds,  ASP Conf. Ser.
  Vol. 101, Astronomical Data Analysis Software and Systems V. p.~17

\bibitem[\protect\citeauthoryear{{Bardeen}, {Press}  \& {Teukolsky}}{{Bardeen}
  et~al.}{1972}]{bar1972}
{Bardeen} J.~M.,  {Press} W.~H.,   {Teukolsky} S.~A.,  1972, \mn@doi [\apj]
  {10.1086/151796}, \href
  {https://ui.adsabs.harvard.edu/abs/1972ApJ...178..347B} {178, 347}

\bibitem[\protect\citeauthoryear{{Boldt}}{{Boldt}}{1987}]{bol1987}
{Boldt} E.,  1987, in {Hewitt} A.,  {Burbidge} G.,   {Fang} L.~Z.,  eds,  IAU
  Symp. Vol. 124, Observational Cosmology. p.~611

\bibitem[\protect\citeauthoryear{{Brenneman} \& {Reynolds}}{{Brenneman} \&
  {Reynolds}}{2006}]{bre2006}
{Brenneman} L.~W.,  {Reynolds} C.~S.,  2006, \mn@doi [\apj] {10.1086/508146},
  \href {http://adsabs.harvard.edu/abs/2006ApJ...652.1028B} {652, 1028}

\bibitem[\protect\citeauthoryear{{Brenneman} et~al.,}{{Brenneman}
  et~al.}{2011}]{bre2011}
{Brenneman} L.~W.,  et~al., 2011, \mn@doi [\apj] {10.1088/0004-637X/736/2/103},
  \href {https://ui.adsabs.harvard.edu/abs/2011ApJ...736..103B} {736, 103}

\bibitem[\protect\citeauthoryear{{Dauser}, {Wilms}, {Reynolds}  \&
  {Brenneman}}{{Dauser} et~al.}{2010}]{dau2010}
{Dauser} T.,  {Wilms} J.,  {Reynolds} C.~S.,   {Brenneman} L.~W.,  2010,
  \mn@doi [\mnras] {10.1111/j.1365-2966.2010.17393.x}, \href
  {http://ads.bao.ac.cn/abs/2010MNRAS.409.1534D} {409, 1534}

\bibitem[\protect\citeauthoryear{{Dauser}, {Garcia}, {Wilms}, {B{\"o}ck},
  {Brenneman}, {Falanga}, {Fukumura}  \& {Reynolds}}{{Dauser}
  et~al.}{2013}]{dau2013}
{Dauser} T.,  {Garcia} J.,  {Wilms} J.,  {B{\"o}ck} M.,  {Brenneman} L.~W.,
  {Falanga} M.,  {Fukumura} K.,   {Reynolds} C.~S.,  2013, \mn@doi [\mnras]
  {10.1093/mnras/sts710}, \href
  {http://adsabs.harvard.edu/abs/2013MNRAS.430.1694D} {430, 1694}

\bibitem[\protect\citeauthoryear{{Dauser}, {Garc{\'{\i}}a}, {Parker}, {Fabian}
  \& {Wilms}}{{Dauser} et~al.}{2014}]{dau2014}
{Dauser} T.,  {Garc{\'{\i}}a} J.,  {Parker} M.~L.,  {Fabian} A.~C.,   {Wilms}
  J.,  2014, \mn@doi [\mnras] {10.1093/mnrasl/slu125}, \href
  {http://ads.bao.ac.cn/abs/2014MNRAS.444L.100D} {444, L100}

\bibitem[\protect\citeauthoryear{{Davis}}{{Davis}}{2001}]{dav2001}
{Davis} J.~E.,  2001, \mn@doi [\apj] {10.1086/323488}, \href
  {http://adsabs.harvard.edu/abs/2001ApJ...562..575D} {562, 575}

\bibitem[\protect\citeauthoryear{{Duro, Refiz} et~al.,}{{Duro, Refiz}
  et~al.}{2016}]{dur2016}
{Duro, Refiz} et~al., 2016, \mn@doi [A&A] {10.1051/0004-6361/201424740}, 589,
  A14

\bibitem[\protect\citeauthoryear{{Ebisawa} et~al.,}{{Ebisawa}
  et~al.}{1994}]{ebi1994}
{Ebisawa} K.,  et~al., 1994, \pasj, \href
  {http://ads.bao.ac.cn/abs/1994PASJ...46..375E} {46, 375}

\bibitem[\protect\citeauthoryear{Fabian \& Wilkins}{Fabian \&
  Wilkins}{2011}]{fab2011}
Fabian A.~C.,  Wilkins D.~R.,  2011, \mn@doi [\mnras]
  {10.1111/j.1365-2966.2011.18458.x}, 414, 1269

\bibitem[\protect\citeauthoryear{{Fabian}, {Rees}, {Stella}  \&
  {White}}{{Fabian} et~al.}{1989}]{fab1989}
{Fabian} A.~C.,  {Rees} M.~J.,  {Stella} L.,   {White} N.~E.,  1989, \mn@doi
  [\mnras] {10.1093/mnras/238.3.729}, \href
  {http://ads.bao.ac.cn/abs/1989MNRAS.238..729F} {238, 729}

\bibitem[\protect\citeauthoryear{{Fabian}, {Iwasawa}, {Reynolds}  \&
  {Young}}{{Fabian} et~al.}{2000}]{fab2000}
{Fabian} A.~C.,  {Iwasawa} K.,  {Reynolds} C.~S.,   {Young} A.~J.,  2000,
  \mn@doi [\pasp] {10.1086/316610}, \href
  {http://ads.bao.ac.cn/abs/2000PASP..112.1145F} {112, 1145}

\bibitem[\protect\citeauthoryear{Fabian et~al.,}{Fabian
  et~al.}{2012}]{fab2012a}
Fabian A.~C.,  et~al., 2012, \mn@doi [\mnras]
  {10.1111/j.1365-2966.2011.19676.x}, 419, 116

\bibitem[\protect\citeauthoryear{Fabian et~al.,}{Fabian et~al.}{2014}]{fab2014}
Fabian A.~C.,  et~al., 2014, \mn@doi [\mnras] {10.1093/mnras/stu1246}, 443,
  1723

\bibitem[\protect\citeauthoryear{Fabian et~al.,}{Fabian et~al.}{2018}]{fab2018}
Fabian A.~C.,  et~al., 2018, \mn@doi [\mnras] {10.1093/mnras/sty836}, 477, 3711

\bibitem[\protect\citeauthoryear{{Ferrigno}, {Bozzo}, {Del Santo}  \&
  {Capitanio}}{{Ferrigno} et~al.}{2012}]{fer2012}
{Ferrigno} C.,  {Bozzo} E.,  {Del Santo} M.,   {Capitanio} F.,  2012, \mn@doi
  [\aap] {10.1051/0004-6361/201118474}, \href
  {https://ui.adsabs.harvard.edu/abs/2012A&A...537L...7F} {537, L7}

\bibitem[\protect\citeauthoryear{{Fukazawa} et~al.,}{{Fukazawa}
  et~al.}{2009}]{fuk2009}
{Fukazawa} Y.,  et~al., 2009, \mn@doi [\pasj] {10.1093/pasj/61.sp1.S17}, \href
  {http://adsabs.harvard.edu/abs/2009PASJ...61S..17F} {61, S17}

\bibitem[\protect\citeauthoryear{{Garc{\'{\i}}a} \& {Kallman}}{{Garc{\'{\i}}a}
  \& {Kallman}}{2010}]{gar2010}
{Garc{\'{\i}}a} J.,  {Kallman} T.~R.,  2010, \mn@doi [\apj]
  {10.1088/0004-637X/718/2/695}, \href
  {http://adsabs.harvard.edu/abs/2010ApJ...718..695G} {718, 695}

\bibitem[\protect\citeauthoryear{{Garc{\'{\i}}a}, {Kallman}  \&
  {Mushotzky}}{{Garc{\'{\i}}a} et~al.}{2011}]{gar2011}
{Garc{\'{\i}}a} J.,  {Kallman} T.~R.,   {Mushotzky} R.~F.,  2011, \mn@doi
  [\apj] {10.1088/0004-637X/731/2/131}, \href
  {http://adsabs.harvard.edu/abs/2011ApJ...731..131G} {731, 131}

\bibitem[\protect\citeauthoryear{{Garc{\'{\i}}a}, {Dauser}, {Reynolds},
  {Kallman}, {McClintock}, {Wilms}  \& {Eikmann}}{{Garc{\'{\i}}a}
  et~al.}{2013}]{gar2013}
{Garc{\'{\i}}a} J.,  {Dauser} T.,  {Reynolds} C.~S.,  {Kallman} T.~R.,
  {McClintock} J.~E.,  {Wilms} J.,   {Eikmann} W.,  2013, \mn@doi [\apj]
  {10.1088/0004-637X/768/2/146}, \href
  {http://ads.bao.ac.cn/abs/2013ApJ...768..146G} {768, 146}

\bibitem[\protect\citeauthoryear{{Garc{\'{\i}}a} et~al.,}{{Garc{\'{\i}}a}
  et~al.}{2014}]{gar2014a}
{Garc{\'{\i}}a} J.,  et~al., 2014, \mn@doi [\apj] {10.1088/0004-637X/782/2/76},
  \href {http://ads.bao.ac.cn/abs/2014ApJ...782...76G} {782, 76}

\bibitem[\protect\citeauthoryear{Garc{\'{\i}}a, Steiner, McClintock, Remillard,
  Grinberg  \& Dauser}{Garc{\'{\i}}a et~al.}{2015}]{gar2015}
Garc{\'{\i}}a J.~A.,  Steiner J.~F.,  McClintock J.~E.,  Remillard R.~A.,
  Grinberg V.,   Dauser T.,  2015, \mn@doi [\apj] {10.1088/0004-637x/813/2/84},
  813, 84

\bibitem[\protect\citeauthoryear{{Garc{\'\i}a}, {Fabian}, {Kallman}, {Dauser},
  {Parker}, {McClintock}, {Steiner}  \& {Wilms}}{{Garc{\'\i}a}
  et~al.}{2016}]{gar2016}
{Garc{\'\i}a} J.~A.,  {Fabian} A.~C.,  {Kallman} T.~R.,  {Dauser} T.,  {Parker}
  M.~L.,  {McClintock} J.~E.,  {Steiner} J.~F.,   {Wilms} J.,  2016, \mn@doi
  [\mnras] {10.1093/mnras/stw1696}, \href
  {https://ui.adsabs.harvard.edu/abs/2016MNRAS.462..751G} {462, 751}

\bibitem[\protect\citeauthoryear{{Garc{\'\i}a}, {Kallman}, {Bautista},
  {Mendoza}, {Deprince}, {Palmeri}  \& {Quinet}}{{Garc{\'\i}a}
  et~al.}{2018a}]{gar2018a}
{Garc{\'\i}a} J.~A.,  {Kallman} T.~R.,  {Bautista} M.,  {Mendoza} C.,
  {Deprince} J.,  {Palmeri} P.,   {Quinet} P.,  2018a, {The Problem of the High
  Iron Abundance in Accretion Disks around Black Holes}

\bibitem[\protect\citeauthoryear{{Garc{\'{\i}}a} et~al.,}{{Garc{\'{\i}}a}
  et~al.}{2018b}]{gar2018}
{Garc{\'{\i}}a} J.~A.,  et~al., 2018b, \mn@doi [\apj]
  {10.3847/1538-4357/aad231}, \href
  {http://ads.bao.ac.cn/abs/2018ApJ...864...25G} {864, 25}

\bibitem[\protect\citeauthoryear{{Jana}, {Debnath}, {Chakrabarti}, {Mondal},
  {Molla}  \& {Chatterjee}}{{Jana} et~al.}{2018}]{jan2018}
{Jana} A.,  {Debnath} D.,  {Chakrabarti} S.~K.,  {Mondal} S.,  {Molla} A.~A.,
  {Chatterjee} D.,  2018, in {Bianchi} M.,  {Jansen} R.~T.,   {Ruffini} R.,
  eds, Fourteenth Marcel Grossmann Meeting - MG14. pp 1038--1043 (\mn@eprint
  {arXiv} {1812.01918}), \mn@doi{10.1142/9789813226609_0058}

\bibitem[\protect\citeauthoryear{{Jiang}, {Walton}, {Fabian}  \&
  {Parker}}{{Jiang} et~al.}{2019}]{jia2019}
{Jiang} J.,  {Walton} D.~J.,  {Fabian} A.~C.,   {Parker} M.~L.,  2019, \mn@doi
  [\mnras] {10.1093/mnras/sty3228}, \href
  {http://ads.bao.ac.cn/abs/2019MNRAS.483.2958J} {483, 2958}

\bibitem[\protect\citeauthoryear{Kammoun, Domček, Svoboda, Dovčiak  \&
  Matt}{Kammoun et~al.}{2019}]{kam2019}
Kammoun E.~S.,  Domček V.,  Svoboda J.,  Dovčiak M.,   Matt G.,  2019,
  \mn@doi [\mnras] {10.1093/mnras/stz408}, 485, 239

\bibitem[\protect\citeauthoryear{{Kennea} et~al.,}{{Kennea}
  et~al.}{2011}]{ken2011}
{Kennea} J.~A.,  et~al., 2011, ATel, \href
  {http://adsabs.harvard.edu/abs/2011ATel.3613....1K} {3613, 1}

\bibitem[\protect\citeauthoryear{{Koyama} et~al.,}{{Koyama}
  et~al.}{2007}]{koy2007}
{Koyama} K.,  et~al., 2007, \mn@doi [\pasj] {10.1093/pasj/59.sp1.S23}, \href
  {http://adsabs.harvard.edu/abs/2007PASJ...59S..23K} {59, 23}

\bibitem[\protect\citeauthoryear{{Laor}}{{Laor}}{1991}]{lao1991}
{Laor} A.,  1991, \mn@doi [\apj] {10.1086/170257}, \href
  {http://ads.bao.ac.cn/abs/1991ApJ...376...90L} {376, 90}

\bibitem[\protect\citeauthoryear{{Lohfink}, {Reynolds}, {Miller}, {Brenneman},
  {Mushotzky}, {Nowak}  \& {Fabian}}{{Lohfink} et~al.}{2012}]{loh2012}
{Lohfink} A.~M.,  {Reynolds} C.~S.,  {Miller} J.~M.,  {Brenneman} L.~W.,
  {Mushotzky} R.~F.,  {Nowak} M.~A.,   {Fabian} A.~C.,  2012, \mn@doi [\apj]
  {10.1088/0004-637X/758/1/67}, \href
  {https://ui.adsabs.harvard.edu/abs/2012ApJ...758...67L} {758, 67}

\bibitem[\protect\citeauthoryear{{L{\'o}pez}, {Jonker}, {Torres}, {Heida},
  {Rau}  \& {Steeghs}}{{L{\'o}pez} et~al.}{2019}]{lop2019}
{L{\'o}pez} K.~M.,  {Jonker} P.~G.,  {Torres} M.~A.~P.,  {Heida} M.,  {Rau} A.,
    {Steeghs} D.,  2019, \mn@doi [\mnras] {10.1093/mnras/sty2793}, \href
  {https://ui.adsabs.harvard.edu/abs/2019MNRAS.482.2149L} {482, 2149}

\bibitem[\protect\citeauthoryear{{Martocchia} \& {Matt}}{{Martocchia} \&
  {Matt}}{1996}]{mar1996}
{Martocchia} A.,  {Matt} G.,  1996, \mn@doi [\mnras] {10.1093/mnras/282.4.L53},
  \href {https://ui.adsabs.harvard.edu/abs/1996MNRAS.282L..53M} {282, L53}

\bibitem[\protect\citeauthoryear{{Matt}, {Perola}  \& {Piro}}{{Matt}
  et~al.}{1991}]{mat1991}
{Matt} G.,  {Perola} G.~C.,   {Piro} L.,  1991, \aap, \href
  {https://ui.adsabs.harvard.edu/abs/1991A&A...247...25M} {247, 25}

\bibitem[\protect\citeauthoryear{{McClintock}, {Shafee}, {Narayan},
  {Remillard}, {Davis}  \& {Li}}{{McClintock} et~al.}{2006}]{mcc2006}
{McClintock} J.~E.,  {Shafee} R.,  {Narayan} R.,  {Remillard} R.~A.,  {Davis}
  S.~W.,   {Li} L.-X.,  2006, \mn@doi [\apj] {10.1086/508457}, \href
  {http://adsabs.harvard.edu/abs/2006ApJ...652..518M} {652, 518}

\bibitem[\protect\citeauthoryear{{Miller-Jones}, {Sivakoff}, {Rupen}  \&
  {Altamirano}}{{Miller-Jones} et~al.}{2011}]{mil2011}
{Miller-Jones} J.~C.~A.,  {Sivakoff} G.~R.,  {Rupen} M.,   {Altamirano} D.,
  2011, ATel, \href {http://adsabs.harvard.edu/abs/2011ATel.3628....1M} {3628,
  1}

\bibitem[\protect\citeauthoryear{{Miller}, {Reynolds}, {Fabian}, {Miniutti}  \&
  {Gallo}}{{Miller} et~al.}{2009}]{mil2009}
{Miller} J.~M.,  {Reynolds} C.~S.,  {Fabian} A.~C.,  {Miniutti} G.,   {Gallo}
  L.~C.,  2009, \mn@doi [\apj] {10.1088/0004-637X/697/1/900}, \href
  {http://adsabs.harvard.edu/abs/2009ApJ...697..900M} {697, 900}

\bibitem[\protect\citeauthoryear{{Miller} et~al.,}{{Miller}
  et~al.}{2013}]{mil2013}
{Miller} J.~M.,  et~al., 2013, \mn@doi [\apj] {10.1088/2041-8205/775/2/L45},
  \href {https://ui.adsabs.harvard.edu/abs/2013ApJ...775L..45M} {775, L45}

\bibitem[\protect\citeauthoryear{{Miniutti} \& {Fabian}}{{Miniutti} \&
  {Fabian}}{2004}]{min2004}
{Miniutti} G.,  {Fabian} A.~C.,  2004, \mn@doi [\mnras]
  {10.1111/j.1365-2966.2004.07611.x}, \href
  {http://ads.bao.ac.cn/abs/2004MNRAS.349.1435M} {349, 1435}

\bibitem[\protect\citeauthoryear{{Mitsuda} et~al.,}{{Mitsuda}
  et~al.}{1984}]{mit1984}
{Mitsuda} K.,  et~al., 1984, \pasj, \href
  {http://ads.bao.ac.cn/abs/1984PASJ...36..741M} {36, 741}

\bibitem[\protect\citeauthoryear{{Nakahira} et~al.,}{{Nakahira}
  et~al.}{2011}]{nak2011}
{Nakahira} S.,  et~al., 2011, ATel, \href
  {http://adsabs.harvard.edu/abs/2011ATel.3626....1N} {3626, 1}

\bibitem[\protect\citeauthoryear{{Negoro} et~al.,}{{Negoro}
  et~al.}{2011}]{neg2011}
{Negoro} H.,  et~al., 2011, ATel, \href
  {http://adsabs.harvard.edu/abs/2011ATel.3611....1N} {3611, 1}

\bibitem[\protect\citeauthoryear{{Novikov} \& {Thorne}}{{Novikov} \&
  {Thorne}}{1973}]{nov1973}
{Novikov} I.~D.,  {Thorne} K.~S.,  1973, in {Dewitt} C.,  {Dewitt} B.~S.,  eds,
  Black Holes (Les Astres Occlus). pp 343--450

\bibitem[\protect\citeauthoryear{{Rau}, {Greiner}  \& {Sudilovsky}}{{Rau}
  et~al.}{2011}]{rau2011}
{Rau} A.,  {Greiner} J.,   {Sudilovsky} V.,  2011, ATel, \href
  {http://adsabs.harvard.edu/abs/2011ATel.3619....1R} {3619, 1}

\bibitem[\protect\citeauthoryear{Reis, Miller, Reynolds, Fabian  \&
  Walton}{Reis et~al.}{2012}]{rei2012}
Reis R.~C.,  Miller J.~M.,  Reynolds M.~T.,  Fabian A.~C.,   Walton D.~J.,
  2012, \mn@doi [\apj] {10.1088/0004-637x/751/1/34}, 751, 34

\bibitem[\protect\citeauthoryear{Reynolds}{Reynolds}{2014}]{rey2013}
Reynolds C.~S.,  2014, \mn@doi [Space Sci. Rev.] {10.1007/s11214-013-0006-6},
  183, 277

\bibitem[\protect\citeauthoryear{{Reynolds} \& {Begelman}}{{Reynolds} \&
  {Begelman}}{1997}]{rey1997}
{Reynolds} C.~S.,  {Begelman} M.~C.,  1997, \mn@doi [\apj] {10.1086/304703},
  \href {http://ads.bao.ac.cn/abs/1997ApJ...488..109R} {488, 109}

\bibitem[\protect\citeauthoryear{Ross \& Fabian}{Ross \&
  Fabian}{2007}]{ros2007}
Ross R.~R.,  Fabian A.~C.,  2007, \mn@doi [\mnras]
  {10.1111/j.1365-2966.2007.12339.x}, 381, 1697

\bibitem[\protect\citeauthoryear{{Russell}, {Soria}, {Motch}, {Pakull},
  {Torres}, {Curran}, {Jonker}  \& {Miller-Jones}}{{Russell}
  et~al.}{2014a}]{rus2014a}
{Russell} T.~D.,  {Soria} R.,  {Motch} C.,  {Pakull} M.~W.,  {Torres} M.~A.~P.,
   {Curran} P.~A.,  {Jonker} P.~G.,   {Miller-Jones} J.~C.~A.,  2014a, \mn@doi
  [\mnras] {10.1093/mnras/stt2480}, \href
  {https://ui.adsabs.harvard.edu/abs/2014MNRAS.439.1381R} {439, 1381}

\bibitem[\protect\citeauthoryear{{Russell}, {Soria}, {Miller-Jones}, {Curran},
  {Markoff}, {Russell}  \& {Sivakoff}}{{Russell} et~al.}{2014b}]{rus2014b}
{Russell} T.~D.,  {Soria} R.,  {Miller-Jones} J.~C.~A.,  {Curran} P.~A.,
  {Markoff} S.,  {Russell} D.~M.,   {Sivakoff} G.~R.,  2014b, \mn@doi [\mnras]
  {10.1093/mnras/stt2498}, \href
  {https://ui.adsabs.harvard.edu/abs/2014MNRAS.439.1390R} {439, 1390}

\bibitem[\protect\citeauthoryear{{Shakura} \& {Sunyaev}}{{Shakura} \&
  {Sunyaev}}{1973}]{sha1973}
{Shakura} N.~I.,  {Sunyaev} R.~A.,  1973, \aap, \href
  {http://cdsads.u-strasbg.fr/abs/1973A%26A....24..337S} {24, 337}

\bibitem[\protect\citeauthoryear{{Strohmayer} \& {Smith}}{{Strohmayer} \&
  {Smith}}{2011}]{str2011}
{Strohmayer} T.~E.,  {Smith} E.~A.,  2011, ATel, \href
  {http://adsabs.harvard.edu/abs/2011ATel.3618....1S} {3618, 1}

\bibitem[\protect\citeauthoryear{{Svoboda}, {Dov{\v c}iak, M.}, {Goosmann, R.
  W.}, {Jethwa, P.}, {Karas, V.}, {Miniutti, G.}  \& {Guainazzi, M.}}{{Svoboda}
  et~al.}{2012}]{svo2012}
{Svoboda} J.,  {Dov{\v c}iak, M.} {Goosmann, R. W.} {Jethwa, P.} {Karas, V.}
  {Miniutti, G.}  {Guainazzi, M.} 2012, \mn@doi [A&A]
  {10.1051/0004-6361/201219701}, 545, A106

\bibitem[\protect\citeauthoryear{{Takahashi} et~al.,}{{Takahashi}
  et~al.}{2007}]{tak2007}
{Takahashi} T.,  et~al., 2007, \mn@doi [\pasj] {10.1093/pasj/59.sp1.S35}, \href
  {http://adsabs.harvard.edu/abs/2007PASJ...59S..35T} {59, 35}

\bibitem[\protect\citeauthoryear{{Tomsick} et~al.,}{{Tomsick}
  et~al.}{2018}]{tom2018}
{Tomsick} J.~A.,  et~al., 2018, \mn@doi [\apj] {10.3847/1538-4357/aaaab1},
  \href {http://ads.bao.ac.cn/abs/2018ApJ...855....3T} {855, 3}

\bibitem[\protect\citeauthoryear{{Tripathi}, {Nampalliwar}, {Abdikamalov},
  {Ayzenberg}, {Bambi}, {Dauser}, {Garc{\'\i}a}  \& {Marinucci}}{{Tripathi}
  et~al.}{2019}]{tri2019}
{Tripathi} A.,  {Nampalliwar} S.,  {Abdikamalov} A.~B.,  {Ayzenberg} D.,
  {Bambi} C.,  {Dauser} T.,  {Garc{\'\i}a} J.~A.,   {Marinucci} A.,  2019,
  \mn@doi [\apj] {10.3847/1538-4357/ab0e7e}, \href
  {https://ui.adsabs.harvard.edu/abs/2019ApJ...875...56T} {875, 56}

\bibitem[\protect\citeauthoryear{{Verner}, {Ferland}, {Korista}  \&
  {Yakovlev}}{{Verner} et~al.}{1996}]{ver1996}
{Verner} D.~A.,  {Ferland} G.~J.,  {Korista} K.~T.,   {Yakovlev} D.~G.,  1996,
  \mn@doi [\apj] {10.1086/177435}, \href
  {http://adsabs.harvard.edu/abs/1996ApJ...465..487V} {465, 487}

\bibitem[\protect\citeauthoryear{{Walton}, {Nardini}, {Fabian}, {Gallo}  \&
  {Reis}}{{Walton} et~al.}{2013}]{wal2013}
{Walton} D.~J.,  {Nardini} E.,  {Fabian} A.~C.,  {Gallo} L.~C.,   {Reis} R.~C.,
   2013, \mn@doi [\mnras] {10.1093/mnras/sts227}, \href
  {https://ui.adsabs.harvard.edu/abs/2013MNRAS.428.2901W} {428, 2901}

\bibitem[\protect\citeauthoryear{{Walton} et~al.,}{{Walton}
  et~al.}{2016}]{wal2016}
{Walton} D.~J.,  et~al., 2016, \mn@doi [\apj] {10.3847/0004-637X/826/1/87},
  \href {https://ui.adsabs.harvard.edu/abs/2016ApJ...826...87W} {826, 87}

\bibitem[\protect\citeauthoryear{{Walton} et~al.,}{{Walton}
  et~al.}{2019}]{wal2019}
{Walton} D.~J.,  et~al., 2019, \mn@doi [\mnras] {10.1093/mnras/stz115}, \href
  {https://ui.adsabs.harvard.edu/abs/2019MNRAS.484.2544W} {484, 2544}

\bibitem[\protect\citeauthoryear{{Wang}, {Ghasemi-Nodehi}, {Guainazzi}  \&
  {Bambi}}{{Wang} et~al.}{2017}]{wan2017}
{Wang} Y.,  {Ghasemi-Nodehi} M.,  {Guainazzi} M.,   {Bambi} C.,  2017, arXiv
  e-prints, \href {https://ui.adsabs.harvard.edu/abs/2017arXiv170307182W} {p.
  arXiv:1703.07182}

\bibitem[\protect\citeauthoryear{{Wilkins} \& {Fabian}}{{Wilkins} \&
  {Fabian}}{2012}]{wil2012b}
{Wilkins} D.~R.,  {Fabian} A.~C.,  2012, \mn@doi [\mnras]
  {10.1111/j.1365-2966.2012.21308.x}, \href
  {https://ui.adsabs.harvard.edu/abs/2012MNRAS.424.1284W} {424, 1284}

\bibitem[\protect\citeauthoryear{Wilkins \& Gallo}{Wilkins \&
  Gallo}{2015}]{wil2015}
Wilkins D.~R.,  Gallo L.~C.,  2015, \mn@doi [\mnras] {10.1093/mnras/stv162},
  449, 129

\bibitem[\protect\citeauthoryear{Wilkins et~al.,}{Wilkins
  et~al.}{2012}]{wil2012a}
Wilkins D.~R.,  et~al., 2012, \mn@doi [\mnras]
  {10.1111/j.1365-2966.2012.21185.x}, 424, 217

\bibitem[\protect\citeauthoryear{{Wilms}, {Allen}  \& {McCray}}{{Wilms}
  et~al.}{2000}]{wil2000}
{Wilms} J.,  {Allen} A.,   {McCray} R.,  2000, \mn@doi [\apj] {10.1086/317016},
  \href {http://ads.bao.ac.cn/abs/2000ApJ...542..914W} {542, 914}

\bibitem[\protect\citeauthoryear{{Xu} et~al.,}{{Xu} et~al.}{2018}]{xu2018}
{Xu} Y.,  et~al., 2018, \mn@doi [\apjl] {10.3847/2041-8213/aaa4b2}, \href
  {http://ads.bao.ac.cn/abs/2018ApJ...852L..34X} {852, L34}

\bibitem[\protect\citeauthoryear{{Yamada} et~al.,}{{Yamada}
  et~al.}{2012}]{yam2012}
{Yamada} S.,  et~al., 2012, \mn@doi [\pasj] {10.1093/pasj/64.3.53}, \href
  {http://ads.nao.ac.jp/abs/2012PASJ...64...53Y} {64, 53}

\bibitem[\protect\citeauthoryear{Young, Ross  \& Fabian}{Young
  et~al.}{1999}]{you1999}
Young A.~J.,  Ross R.~R.,   Fabian A.~C.,  1999, \mn@doi [\mnras]
  {10.1046/j.1365-8711.1999.02528.x}, 306, 461

\bibitem[\protect\citeauthoryear{{Zdziarski}, {Johnson}  \&
  {Magdziarz}}{{Zdziarski} et~al.}{1996}]{zdz1996}
{Zdziarski} A.~A.,  {Johnson} W.~N.,   {Magdziarz} P.,  1996, \mn@doi [\mnras]
  {10.1093/mnras/283.1.193}, \href
  {http://adsabs.harvard.edu/abs/1996MNRAS.283..193Z} {283, 193}

\bibitem[\protect\citeauthoryear{{{\.Z}ycki}, {Done}  \& {Smith}}{{{\.Z}ycki}
  et~al.}{1999}]{zyc1999}
{{\.Z}ycki} P.~T.,  {Done} C.,   {Smith} D.~A.,  1999, \mn@doi [\mnras]
  {10.1046/j.1365-8711.1999.02885.x}, \href
  {http://adsabs.harvard.edu/abs/1999MNRAS.309..561Z} {309, 561}

\makeatother
\end{thebibliography}


\bsp	
\label{lastpage}
\end{document}